\documentclass[showpacs,twocolumn,prl,aps,a4]{revtex4}

\usepackage{graphicx}% Include figure files
\usepackage{epsfig}
\usepackage{dcolumn}% Align table columns on decimal point
\usepackage{bm}% bold math

\begin{document}

\title[Quantum dynamics of a resonator]{Quantum dynamics of a resonator driven by a superconducting single-electron
transistor: a solid-state analogue of the micromaser}%
\author{D.A. Rodrigues, J. Imbers and A.D. Armour}%
\affiliation{School of Physics and Astronomy, University of
Nottingham, Nottingham NG7 2RD, United Kingdom}%

% ----------------------------------------------------------------
\begin{abstract}
We investigate the behavior of a quantum resonator coupled to a
superconducting single-electron transistor tuned to the Josephson
quasiparticle resonance and show that the dynamics is similar in
many ways to that found in a micromaser. Coupling to the SSET can
drive the resonator into non-classical states of self-sustained
oscillation via either continuous or discontinuous transitions.
Increasing the coupling further leads to a sequence of transitions
and regions of multistability.
\end{abstract}


\maketitle
% ----------------------------------------------------------------

Systems where a mesoscopic conductor such as a single-electron
transistor is coupled to a nanomechanical resonator have been
studied intensively because the current through the conductor can be
extremely sensitive to the motion of the resonator and hence may be
used to monitor its position with almost quantum-limited
precision\,\cite{Blencowe,SET-expt,SSET-expt,SSET1}. Furthermore,
where either the coupling between the electrons and the resonator is
non-linear\,\cite{shuttle} or the electronic transport occurs via a
resonance\,\cite{SSET1}, dynamic instabilities in the resonator can
occur leading to self-sustained oscillations. The way a
nanomechanical resonator can be driven into states of finite
amplitude oscillation by successive interactions with a current of
electrons in a conductor parallels the behavior of quantum optical
systems, such as the micromaser, in which an electromagnetic cavity
is pumped by interactions with a steady stream of individual
two-level atoms\,\cite{micromaser_review}. This contrasts with a
standard laser (a nanomechanical version of which was envisioned in
\cite{bargatin}) where an oscillator interacts simultaneously with
many two-level systems.

In a superconducting single-electron transistor (SSET) transport
can occur via resonant processes involving both coherent motion of
Cooper pairs and incoherent quasiparticle tunneling, the simplest
of which is the Josephson quasiparticle (JQP)
resonance\,\cite{Choi}. In the vicinity of a JQP resonance, the
dynamics of a resonator coupled linearly to the SSET is very
sensitive to the bias point\,\cite{SSET1,SSET2,SSET-expt}. For
bias points on one side of the resonance, the SSET acts on the
resonator like a thermal bath and its current can monitor the
position of the resonator with exquisite sensitivity. In contrast,
biasing on the opposite side of the JQP resonance can drive the
resonator into states of self-sustained oscillation\,\cite{SSET1}.

In this Letter we explore the quantum dynamics of a resonator
coupled to a SSET and show that it is analogous to that of a
micromaser. Less noisy than a laser, a
micromaser\,\cite{micromaser_review,micromaser} can generate
number-squeezed states of the cavity and exhibits not a single
threshold transition, but a series of transitions between
different dynamical states. Although the SSET-resonator system and
micromaser differ in the details of the interactions between their
respective sub-components, we find a number of important
similarities in their dynamics, many of which first arise when the
resonator is sufficiently fast to match the time-scale of the
electrical transport. Previous theoretical studies of this system
have concentrated on the limit of a slow
resonator\,\cite{SSET1,SSET2} as it is only this limit which has
so far been explored in experiments on nanomechanical resonators
coupled to a SSET\,\cite{SET-expt,SSET-expt}. The much faster
resonator speeds which we also consider here might be achieved by
making a smaller\,\cite{GHZ} mechanical resonator. However, it
should also be possible to use a superconducting resonator (e.g.\
 a stripline resonator\,\cite{SCR}) for which much higher
frequencies are practical.

\begin{figure}[t]
\center{ \epsfig{file=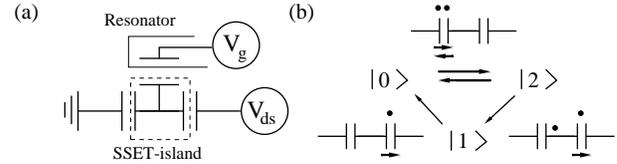,width=8.0cm}
} \caption{(a) Schematic diagram of the SSET-resonator system. The
SSET consists of a superconducting island linked by tunnel junctions
to superconducting leads, across which a voltage $V_{ds}$ is
applied. (b) Charge processes involved in the JQP current cycle
(details are given in the text).} \label{fig:schema}
\end{figure}

The SSET-resonator system we consider is shown schematically in
Fig.\ \ref{fig:schema}a. A mechanical resonator acts as a
voltage-gate with a position dependent capacitance and the coupling
is controlled by varying the voltage applied to it (an analogous,
though fixed, electrostatic coupling arises for a superconducting
stripline resonator\,\cite{SCR}). The SSET is assumed to be biased
close to a JQP resonance\,\cite{Choi} where only three charge states
of the SSET island are relevant. Current flows in a cycle (Fig.\
\ref{fig:schema}b): coherent Josephson tunneling between the left
lead and the island produces a superposition of island charge-states
$|0\rangle$ and $|2\rangle$, this is then followed by two
quasiparticle tunneling events between the island and the right lead
which take the SSET island from state $|2\rangle$ to state
$|1\rangle$ and finally back to state $|0\rangle$. The resonator is
modeled as a single-mode quantum harmonic oscillator with mass $m$
and angular frequency $\omega$. For a weak electrostatic interaction
between the resonator and the SSET, the coupling between the
resonator position and the number of excess charges on the SSET
island is linear\,\cite{SSET1,SSET2}. The SSET-resonator system can
therefore be described by the Hamiltonian
\begin{eqnarray}
H&=&\Delta E |2\rangle \langle 2|
-\frac{E_J}{2}\left(|0\rangle\langle 2|+|2\rangle\langle
0|\right)+\hbar\omega a^{\dagger}a\nonumber \\&&+ x_s(\hbar\omega^3
m/2)^{1/2} (a^{\dagger}+a)\left(|1\rangle\langle
1|+2|2\rangle\langle 2|\right), \label{Ham}
\end{eqnarray}
where $a$ is the annihilation operator for the resonator, $\Delta
E$ is the electrostatic energy difference between the island
states $|0\rangle$ and $|2\rangle$, $x_s$ is the displacement in
the equilibrium position of the resonator when one electronic
charge is added to the SSET island\cite{SSET2} and $E_J$ the
Josephson energy
of the superconductors. %\,\cite{footnote_1}.
 The system can be
considered to be in the weak-coupling limit when the voltage
applied across the SSET is much larger than the coupling energy,
$\kappa=m\omega^2x_s^2/eV_{ds}\ll 1$\,\cite{SSET2}.

The system evolves coherently under the action of the Hamiltonian,
but dissipation arises from two sources: quasiparticle tunneling in
the SSET and the resonator's surroundings. Thus the master equation
for the SSET-resonator density matrix, $\rho(t)$, derived using the
Born-Markov approach\cite{Choi}, takes the form
\begin{equation}
\dot{\rho}={\mathcal
L}\rho=-\frac{i}{\hbar}[H,\rho]+\mathcal{L}_{leads}\rho+\mathcal{L}_{damping}\rho,
\label{master}
\end{equation}
where the dissipative terms arising from the resonator's
surroundings and from quasiparticle tunneling are described by the
terms $\mathcal{L}_{damping}$ and $\mathcal{L}_{leads}$
respectively. The resonator's surroundings are assumed to act like a
thermal bath which we describe using a Liouvillian\,\cite{Flindt},
\begin{eqnarray*}
\mathcal{L}_{damping}\rho&=&-\frac{\gamma_{ext}}{2}(\overline{n}+1)\left(
a^{\dagger}a\rho+\rho a^{\dagger}a-2a\rho a^{\dagger}\right)
 \\ &&-\frac{\gamma_{ext}}{2}\overline{n}\left(
aa^{\dagger}\rho+\rho aa^{\dagger}-2a^{\dagger}\rho a\right),
\nonumber
\end{eqnarray*}
which is guaranteed to preserve the positivity of the density
matrix. The temperature of the external bath is parameterised by the
average number of quanta, $\overline{n}$, that the resonator would
have were it in thermal equilibrium with the bath and the
resonator-bath coupling is given by a damping rate $\gamma_{ext}$.
The tunneling of quasiparticles from the island to the left lead is
described by\,\cite{Choi}
\begin{eqnarray*}
\mathcal{L}_{leads}\rho&=&-\frac{\Gamma}{2}\left[\left\{|2\rangle\langle
2| +|1\rangle\langle
1|,\rho\right\}_+\right.\\&&\left.-2\left(|1\rangle\langle
2|+|0\rangle\langle 1|\right)\rho\left(|2\rangle\langle
1|+|1\rangle\langle 0|\right)\right],\nonumber
\end{eqnarray*}
where $\Gamma$ is the quasiparticle tunneling rate. This is a
simplified expression in which we have neglected differences
between the quasiparticle rates for the two processes, their
variation with bias point and temperature
dependence\,\cite{SSET1,Choi,SSET2}. We have also neglected the
dependence of the quasiparticle tunneling rates on the resonator
position as this is much less important than the coherent
electro-mechanical coupling in the Hamiltonian (Eq.\
\ref{Ham})\,\cite{SSET1,SSET2,footnote_2}. These simplifications
allow us to capture the essential phenomenology of the system
using a relatively compact model. Furthermore, simplified in this
way our model of the SSET-resonator system close to the JQP
resonance is dual to another system, that of a double quantum dot
gated by a resonator\,\cite{BL}.

The value of $\Delta E$ determines the detuning of the Cooper pairs
from resonance and can be changed continuously by changing the
applied gate voltage (this can be done independently of the coupling
provided an additional, fixed, gate is used\,\cite{SSET-expt}). The
sign of $\Delta E$ determines the direction of (average) energy flow
between the resonator and the SSET. For $\Delta E< 0$ the state
$|0\rangle$ has more energy than $|2\rangle$, hence when a Cooper
pair tunnels onto the island it can pass some of its energy to the
resonator (before quasiparticle decays occur), but when $\Delta E>
0$ the situation is reversed and hence the Cooper pair can absorb
energy from the resonator. This means that when $\Delta E>0$ the
SSET damps the motion of the resonator, though because of the
stochastic nature of the current, the resonator settles into a
thermal-like steady state, not its ground state\,\cite{SSET1,SSET2}.
However, when $\Delta E<0$ the transfer of successive Cooper pairs
passing from lead to island can pump the resonator and drive its
oscillation.

\begin{figure}[t]
\center{ \epsfig{file=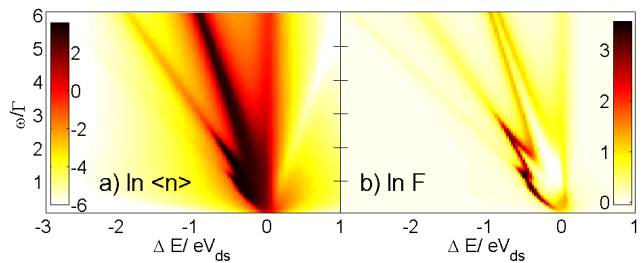, width=8.5cm} }
\caption{(Color online) Steady state properties of the resonator
as a function of $\Delta E$, and resonator speed, $\omega/\Gamma$:
(a) average occupation number of the resonator, $\langle
n\rangle$; (b) Fano factor of the resonator, both plotted on a
natural-log scale, with $\kappa=0.01$,
$\gamma_{ext}/\Gamma=0.002$, and $\overline{n}=0$.} \label{fig:2}
\end{figure}

In order to explore the behavior of the system over a range of
resonator speeds, from slow ($\omega/\Gamma\ll 1$) to fast
($\omega/\Gamma\gg 1$) we solve the master equation using a
numerical method. There are a number of methods which can be
employed to calculate the steady state of a master equation. We have
made use of the numerical routines implemented in the Quantum Optics
Toolbox\,\cite{qotoolbox} to calculate the steady state of the
density matrix by evaluating the eigenfunction corresponding to the
zero eigenvalue of the Liouvillian in Eq.\ \ref{master} (written in
matrix form). Our method necessitates a truncation of the
resonator's Hilbert space, which because of the need to ensure
convergence, effectively translates into a lower limit on the value
of the external damping we consider (we used a Fock state
representation with up to 70 states).

\begin{figure}[t]
\center{ \epsfig{file=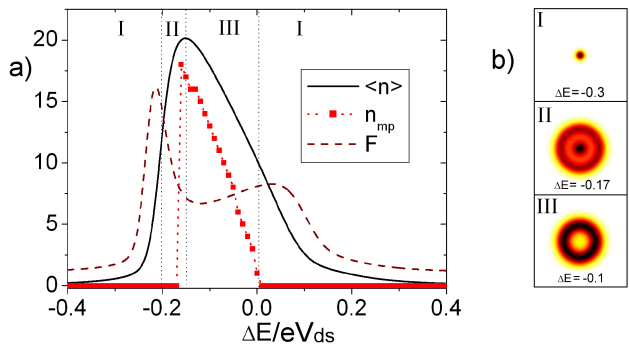, width=8.5cm}}\caption{(Color
online)(a) Average occupation number and Fano factor of the
resonator together with the most probable number state in the
density matrix, $n_{mp}$, for $\omega/\Gamma=0.1$,
$\gamma_{ext}/\Gamma=0.0005$, $\kappa=0.02$ and $\overline{n}=0$.
(b) Examples of the resonator Wigner functions: I fixed point
state, II bistable and III limit cycle. The regions in (a)
corresponding to the three basic Wigner function topologies are
also marked as I (fixed point), II (bistable) and III (limit
cycle).} \label{fig:wigner}
\end{figure}

The effect on the steady state of the resonator of varying the
detuning from resonance, $\Delta E$, for a range of resonators
speeds, $\omega/\Gamma$, is shown in Fig.\ 2 where the average
number of resonator quanta, $\langle n\rangle$ (with
$n=a^{\dagger}a$), and the Fano factor, $F=(\langle
n^2\rangle-\langle n\rangle ^2)/\langle n\rangle$ are plotted. We
assume (throughout) SSET parameters\,\cite{SSET2}
$\Gamma=V_{ds}/eR_J$, $R_J=h/e^2$ and $E_J=hV_{ds}/(16eR_J)$,
where $R_J$ is the SSET junction resistance.

A more complete understanding of the resonator state is obtained
from the Wigner transform of the reduced steady state density
matrix of the resonator\,\cite{shuttle}. Although the Wigner
function of the resonator is different in detail for every set of
parameters in Fig.\ 2, it turns out that because of the weak
SSET-resonator coupling there are just three different topologies
which arise: a state in which the resonator fluctuates about a
fixed point, a limit cycle state in which the resonator undergoes
finite amplitude oscillations and a bistable state in which both
the fixed point and limit cycle type states coexist (illustrated
in Fig.\ \ref{fig:wigner}).

The behavior of the resonator is simplest in the limits of a very
fast ($\omega/\Gamma\gg 1$) or very slow ($\omega/\Gamma\ll 1$)
resonator, where a wide separation of the SSET and resonator
time-scales limits their mutual interaction. In contrast, when
$\omega/\Gamma\sim 1$ the SSET and resonator interact most
strongly and hence it is not surprising that it is in this regime
that the most interesting features of the coupled dynamics first
begin to emerge\,\cite{dar}.

For $\omega/\Gamma\ge 1$,  Fig.\ 2 shows that the transfer of
energy between SSET and resonator is increasingly concentrated
around a series of lines. These lines are points where $\Delta
E=\pm j \hbar \omega$ with $j$ an integer (the current through the
SSET also shows peaks at these values). When $\Delta E<0$, the
resonator absorbs quanta of energy from the SSET leading to an
enhancement of $\langle n\rangle$ and $F$. Because we are working
in the weak coupling regime, the strongest effect occurs for
$\Delta E=-\hbar\omega$, with progressively weaker signatures at
$\Delta E=-2\hbar\omega$ and $- 3\hbar \omega$, as the exchange of
multiple quanta result from higher order processes. A dip in
$\langle n\rangle$ for $\Delta E=+\hbar \omega$ is also visible
though it becomes less pronounced as $\omega/\Gamma$ is increased.
This dip occurs, despite the fact that we have assumed an external
bath at zero temperature, as the noise arising from coupling to
the SSET excites the resonator out of its ground state even for
$\Delta E>0$. These results are consistent with calculations for a
resonator coupled to a double quantum dot in this
regime\,\cite{BL}.

In the limit of a slow resonator, $\omega/\Gamma\ll 1$, variation
of $\Delta E$ leads to two distinct transitions in the state of
the resonator as is shown in Fig.\ \ref{fig:wigner}. The
transitions become sharper as the ratio of the SSET current to the
relaxation rate $\gamma_{ext}$ increases, which suggests that they
can be thought of as non-equilibrium phase transitions in a
`thermodynamic limit' where this ratio
diverges\,\cite{micromaser,Haken}. The most probable number state
in the density matrix, $n_{mp}$, provides a convenient order
parameter of the system. For $\Delta E/eV_{ds} \gtrsim 0.4$, the
resonator is very close to being in a thermal state and hence
there is a single peak in the Wigner function, but as $\Delta E$
is decreased and approaches zero the resonator state broadens
(deviating strongly from a thermal state), until for $\Delta
E\simeq 0$ a limit cycle begins to appear. The transition to a
limit cycle is continuous as the radius of the cycle (and
$n_{mp}$) grows steadily from zero as $\Delta E$ is decreased.
However, eventually (as $\Delta E$ becomes sufficiently negative)
the system passes through a region of bistability where both the
limit cycle and a central peak co-exist in the Wigner function
before the limit cycle disappears entirely and only the central
peak remains. This behavior implies a discontinuous transition and
is marked by a jump in $n_{mp}$.

For a very slow resonator, the current flowing through the SSET
decays monotonically to zero far from the center of the JQP
resonance. This explains why the limit cycle eventually disappears
as $\Delta E$ is decreased: the energy per unit time transferred
to the resonator is proportional to the current and as this
decays, the system will eventually be stabilized by the external
damping.

A striking feature of the intermediate regime ($\omega/\Gamma\sim
1$) in Fig.\ 2 is the drop in the Fano factor of the resonator
which occurs in the limit cycle state. Remarkably, there is a
region where the Fano factor falls below unity, implying that the
resonator is driven into a number squeezed (i.e.\ non-classical)
steady state. We attribute this squeezing to non-linear damping
induced by the interaction with the SSET\,\cite{SSET1}. Squeezing
of this kind is a characteristic of the
micromaser\,\cite{micromaser_review,micromaser}, but never occurs
in the (more noisy) conventional laser\,\cite{Haken}. As with the
micromaser, we find that increasing the noise in the system (by
increasing the temperature of the bath) rapidly washes out the
squeezing effect.

\begin{figure}[t]
\center{ \epsfig{file=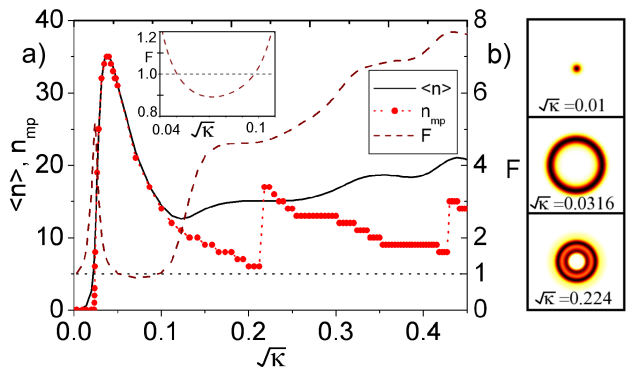, width=8.4cm}
}\caption{(Color online) Changes in the resonator state (a) and
corresponding Wigner functions (b) as the coupling $\kappa$ is
increased for $\Delta E/eV_{ds}=-0.1$, $\omega/\Gamma=1$,
$\gamma_{ext}/\Gamma=0.001$ and $\overline{n}=0$. After an initial
continuous transition ($\kappa \simeq 0.025^2$), the system
undergoes discontinuous transitions at $\kappa \simeq 0.22^2$ and
$\kappa \simeq 0.44^2$. The inset in (a) shows the region where
the Fano factor drops below unity. From top to bottom, the Wigner
functions show: a fixed point state, a single limit cycle and two
metastable limit cycles.} \label{fig:ksweep}
\end{figure}

The regime where $\omega/\Gamma\sim 1$ is also where further
transitions beyond a simple limit cycle state first occur as the
SSET-resonator coupling is made stronger. A hallmark of the
micromaser is the sequence of transitions which occur as the
atom-cavity coupling is increased\,\cite{micromaser}. We find that
the resonator undergoes a very similar sequence of transitions as
the coupling is increased at fixed $\Delta E$($<0$), the first of
which is continuous (for $|\Delta E|$ sufficiently small) and the
rest discontinuous. An example of this behavior is shown in Fig.\
\ref{fig:ksweep}. As the coupling is increased from zero the
resonator first undergoes a continuous transition between fixed
point and limit cycle states. Within the limit cycle state $F$
drops below unity, but as $\kappa$ increases above $0.01$, $F$
grows sharply as a second metastable limit-cycle state begins to
appear. For $\kappa \gtrsim 0.04$, the Wigner functions become
more complicated, showing further metastable limit-cycles together
with regions where parts of the Wigner function have negative
values (indicating non-classicality of the resonator state). The
relative weights of the different limit cycle states change with
$\kappa$, leading to discontinuous transitions signalled by jumps
in $n_{mp}$ as the most probable state of the resonator changes.
In Fig.\ \ref{fig:ksweep} only the first (continuous) transition
shows sharp features in $\langle n\rangle$ and $F$ as we are far
from the `thermodynamic limit'\,\cite{micromaser} (we have chosen
a relatively large $\gamma_{ext}$ to ensure that our truncation of
the oscillator state space remains valid).

 In conclusion, we have
analyzed the dynamics of a resonator coupled to a SSET near the JQP
resonance. The SSET-resonator system has a rich dynamics with many
features in common with the micromaser including continuous and
discontinuous transitions between resonator states and intrinsically
quantum features such as non-classical steady-states. In practice,
the resonator dynamics could be inferred from signatures in the
current and current noise\,\cite{cnoise} of the SSET or, for a
superconducting resonator, by coupling it to a transmission
line\,\cite{SCR}. The effects described are strongest when
$\omega/\Gamma\sim 1$ and will require minimal thermal excitation
($\overline{n}\lesssim 1$) for experimental observation of the
non-classical features, implying high resonator frequencies.
%
%which may be achieved in experiment either by using high resonator
%frequencies or by engineering a SSET with low quasiparticle rates.
%However, we expect that for the quantum features of the dynamics
%to be visible, the thermal excitation of the resonator will need
%to be minimal (i.e. $\overline{n}\lesssim 1$).
Similarities with the micromaser suggest the possibility of
observing critical slowing down, hysteresis and quantum jumps in
the resonator dynamics\,\cite{micromaser_review}.

After submission of this paper, \cite{ashlaser} was published
which explores analogies between nano-electromechanical systems
and laser physics in the slow-resonator limit. We thank T.\
Brandes, M.\ Blencowe and T.\ Harvey for useful discussions and
acknowledge funding from the EPSRC.

% ----------------------------------------------------------------
%\bibliographystyle{amsplain}
%\bibliography{}

\begin{thebibliography}{99}
\bibitem{footnote_2} The numerical method can also be used to
explore the behavior of a more accurate (though less general)
model of the SSET resonator system. Indeed, we have checked that
including the position dependence of the quasiparticle rates does
not qualitatively alter the results we present.
\end{thebibliography}

\end{document}